\documentclass[mathleft
]{an}
\usepackage{graphicx}
\usepackage{times}

\def\ltsima{$\; \buildrel < \over \sim \;$}
\def\simlt{\lower.5ex\hbox{\ltsima}}
\def\gtsima{$\; \buildrel > \over \sim \;$}
\def\simgt{\lower.5ex\hbox{\gtsima}}
\def\cgs{{erg cm$^{-2}$ s$^{-1}$}}
\def\ergs{{erg s$^{-1}$}}
\def\cm2{{cm$^{-2}$}}

\def\lum{{$L_{\rm X}$}}

\def\p1{{Paper I}}

\def\xmm{{\em XMM--Newton}}
\def\chandra{{\em Chandra}}

\def\xmm{{\em XMM--Newton}}

\def\nh{{N$_{\rm H}$}}

\def\xray{{X--ray}}
\def\f14{{10$^{-14}$}}
\def\f13{{10$^{-13}$}}
\def\f12{{10$^{-12}$}}
\def\f11{{10$^{-11}$}}

\def\4u{{4U~1344$-$60}}

\def\lmir{{$L_{\rm 6\mu m}$}}

\def\nus{{\em NuSTAR}}

\def\aap{A\&A}

\begin{document}

\Pagespan{1}{6}
\Yearpublication{}%
\Yearsubmission{}%
\Month{}%
\Volume{}%
\Issue{}%

\title{X-ray selection of Compton Thick AGN at high redshift}

\author{G. Lanzuisi\inst{1,2}\fnmsep\thanks{Corresponding author:
  \email{giorgio.lanzuisi2@unibo.it}\newline}
}
\titlerunning{X-ray selection of CT AGN at high z}
\authorrunning{G. Lanzuisi}
\institute{
Dipartimento di Fisica e Astronomia, Universit\'a degli studi di Bologna, Via Ranzani 1, 40127, Bologna, Italy
\and 
INAF-Osservatorio Astronomico di Bologna, Via Ranzani 1, 40127, Bologna, Italy
}

\received{2016 Sep 23}
\accepted{2016 Oct 24}
\publonline{later}

\keywords{galaxies: active -- X-rays: galaxies}

\abstract{%
  Compton Thick (CT) AGN are a key ingredient of Cosmic X-ray Background (CXB) synthesis models, 
  but are still an elusive component of the AGN population beyond the local Universe.
  Multiwavelength surveys are the only way to find them at $z\simgt0.1$,
  and a deep X-ray coverage is crucial in order to clearly identify them among 
  star forming galaxies.   As an example, the deep and wide COSMOS survey allowed us to select a total of 34
  CT sources. This number is computed from the 64 nominal CT candidates, 
  each counted for its \nh\ probability distribution function.
  For each of these sources, rich multiwavelength information is available, 
  and is used to confirm their obscured nature, by comparing the expected AGN luminosity from spectral energy distribution fitting,
  with the absorption-corrected X-ray luminosity.
  While \chandra\ is more efficient, for a given exposure, in detecting CT candidates in current surveys (by a factor $\sim2$),
  deep \xmm\ pointings of bright sources are vital to fully characterize their properties: 
  \nh\ distribution above $10^{25}$ cm$^{-2}$, reflection intensity etc., all crucial parameters of CXB models.
  Since luminous CT AGN at high redshift are extremely rare, the 
  future of CT studies at high redshift will have to rely on the large area 
  surveys currently underway, such as XMM-XXL and Stripe82, 
  and will then require dedicated follow-up with \xmm, while waiting for the advent of the ESA mission {\it Athena}.}

\maketitle

\section{Introduction}

We know, since the late 90's, that a large fraction (up to 30\%) of local AGN
are obscured by large amounts of gas and dust (e.g. Risaliti et al. 1999), 
above the Compton Thick\footnote{At these high column densities the obscuration is mainly due to 
Compton-scattering, rather than photoelectric absorption.} threshold (CT, \nh$\geq\sigma_T^{-1}\sim1.5\times10^{24}$ cm$^{-2}$).
A similar fraction of CT AGN is required in most Cosmic X-ray Background (CXB) synthesis models
(e.g. Comastri et al. 1995, Gilli et al. 2007):
their flat spectrum is needed in order to reproduce the hump at 20-30 keV observed in the CXB (e.g. Ballantyne et al. 2011).
However, the value of the CT fraction that one can derive is largely uncertain (Treister et al. 2009, 
Ueda et al. 2014) due to degeneracies between several model 
parameters, i.e. primary continuum photon index, reflection fraction, \nh\ distribution above $10^{24}$ cm$^{-2}$, and
high energy cut-off (see e.g. Akylas et al. 2012).

A population of even more deeply obscured sources (\nh$\geq10^{25}$ cm$^{-2}$) 
may be required (Comastri et al. 2015) 
in order to reconcile new estimates of the BH mass function in the local 
Universe (Kormendy \& Ho 2015) with that inferred by integrating the luminosity function of 
observed AGN via the continuity equation (e.g. Soltan 1982; Marconi et al. 2004).
These heavily obscured sources, however, will not contribute much to the CXB, since even the highest energy 
X-rays they produce are blocked by the obscuring material.

Despite their expected large number, CT AGN are very difficult to identify beyond the local Universe, resulting
in a small/negligible fraction of CT AGN blindly identified in deep X-ray surveys 
(e.g. Tozzi et al. 2006, Lanzuisi et al. 2013, Marchesi et al. 2016). Even \nus,
sensitive above 10 keV, was able to put only upper limits to the fraction of CT AGN 
at $z=0.5-1$ (Alexander et al. 2013).
Therefore, several multiwavelength techniques have been developed in the past decade, 
based on known CT AGN broad band properties, to pre-select CT AGN beyond the local Universe. 
IR spectral features, IR colors or mid-IR vs. optical or X-ray flux ratios can be used to 
select red, dusty sources at $z\sim2$
(e.g. Lacy et al. 2004, Martinez-Sansigre et al. 2005, Houck et al. 2005).
High ionization optical emission lines from the narrow line region, such as [OIII] and [NeV] 
can be used to select sources with a deficit in the 
observed X-ray emission, that can be ascribed to strong obscuration of the nucleus 
(e.g. Vignali et al. 2006, Gilli et al. 2010, Mignoli et al. 2013). 

In all these cases, however, the X-ray information (either detections or staking) is crucial in order to  
unambiguously identify a fraction of CT AGN among inactive galaxies with similar properties
(e.g. Fiore et al. 2008, Lanzuisi et al. 2009, Georgantopoulos et al. 2011, Vignali et al. 2014).
X-rays are indeed able to provide the smoking gun of the CT nature of these sources,
thanks to the unique spectral signatures observable, i.e. the flat continuum and the strong Fe K emission line at 6.4 keV.
Furthermore, above \lum\ $\sim 10^{42}$ \ergs, the contamination by star-forming galaxies in the 2-10 keV 
is limited.
Finally, X-ray spectroscopy is favored by the redshift effect: going at high redshift, the Compton hump at 20-30 keV
becomes observable by \chandra\ and \xmm\, and the Fe k$\alpha$ line moves toward lower energies, where the effective area
of current telescopes is larger.

There are, however, important caveats to take into account:
i) The fraction of CT AGN steeply rises from $\sim0$ to the intrinsic value (e.g. 0.3-0.4)
only below a certain flux (e.g. $F<<10^{-15}$ \cgs\ in the 2-10 keV band)
and therefore it is mandatory to reach deep sensitivities, in order to collect sizable samples of CT AGN.
ii) CT AGN are a factor 30-50 fainter than normal AGN below 10 keV rest frame, 
requiring long exposures to collect even few tens of X-ray counts per source. 
iii) The transition between Compton-thin and Compton-thick is smooth, and the spectra of sources 
just above/below the CT threshold are very similar (Murphy \& Yaqoob 2009),
requiring a tailored analysis (see. e.g. Buchner et al. 2015) and possibly
the use of the full \nh\ probability distribution function (PDF) 
when counting/selecting CT AGN,
in order to avoid misclassification in one direction or the other.

For all these reasons, even in the deepest X-ray fields, different analysis of the same data sets
(e.g. Tozzi et al. 2006, Brightman \& Ueda 2012, Georgantopoulos et al. 2013)
give different results, not always in agreement (see Castell\'o-Mor et al. 2013).

Here we present results from the analysis of the large COSMOS-Legacy catalog, that allowed us to select
64 CT AGN candidates among the $\sim4000$ point-like sources.
For each candidate, the full PDF of the \nh\ is taken into account,
to weight each source by its probability of being CT.
The LogN-LogS of CT AGN in three redshift bins (up to $z=3.5$) is also presented.

\section{Sample Selection}

\subsection{The COSMOS survey} 

The 2 deg$^2$ area of the {\it HST} COSMOS Treasury program 
is centered at 10:00:28.6, +02:12:21.0 (Scoville et al. 2007).
The field has a unique deep and wide multi-wavelength coverage, 
from the optical band ({\it Hubble, Subaru, VLT}
and other ground-based telescopes), to the infrared ({\it Spitzer, Herschel}),
\xray\ (\xmm, \chandra\ and \nus) and radio bands.
Large dedicated ground-based spectroscopy programs in the optical 
with all the major optical telescopes have been completed.
Very accurate photometric redshifts are available for both the galaxy population (Ilbert et al. 2009) 
and the AGN population (Salvato et al. 2011).

The COSMOS field has been observed with \xmm\ for a total of $\sim1.5$ Ms at a rather homogeneous depth of
$\sim$60 ks over $\sim2$ deg$^2$ (Hasinger et al. 2007, Cappelluti et al. 2009),
and by \chandra\ with a deeper observations of $\sim160$ ks: the central deg$^2$ was observed 
in 2006-2007 (Elvis et al. 2009, Civano et al. 2012) for a total of 1.8 Ms, while addition 1.2 deg$^2$ were covered
recently (2013-2014) by the Chandra COSMOS-Legacy survey, for a total of 2.8 Ms (Civano et al. 2016, Marchesi et al. 2016a).

The \chandra\ catalog used in this work includes 4016 point-like sources (from a total of 4.6 Ms of \chandra\ observation)   
detected in at least one of the following
three bands: full (F; 0.5-7 keV), soft (S; 0.5-2 keV) and
hard (H; 2-7 keV). Each source was detected in at least
one band with probability of being a spurious detection P$<2\times10^{-5}$. 

\subsection{X-ray spectral analysis}

1949 sources in the COSMOS-Legacy catalog have more than 30 counts (with average $\sim145$ and median $\sim68$ counts).
This threshold allows to derive basic spectral properties (\nh, \lum, see Lanzuisi et al. 2013).
The result of a systematic spectral analysis of all these sources is presented in Marchesi et al. (2016).
The procedure described there, however, is not optimized to look for CT AGN, since at such high column densities,
the simple power-law plus photoelectric absorption model does not reproduce any more the physical processes involved.

We reanalyzed all the available spectra using the BNTorus model (Brightman \& Nandra 2011), specifically
developed to model obscuration at CT regimes. The geometry of the obscuring material is a spherical torus 
which is essentially a sphere with a bi-conical aperture.
We used the model with fixed photon index $\Gamma=1.9$, and with fixed torus half opening angle ($60^\circ$) 
and inclination angle ($80^\circ$). 
The model is therefore very simple, as it uses a single geometry and does not 
allow for any variation between the primary continuum 
and the reflection/line component, or for different value of \nh\ between the absorber and the 
reflector\footnote{as e.g. MYTorus in the decoupled version is able to do (Murphy \& Yaqoob 2009).}.
However this choice is forced by the very limited number of counts available for each source (65\% of the final 
CT sample has less than 50 net counts). 
In addition to the BNTorus model, we used a secondary powerlaw, with the photon index fixed to 1.9,
to model the emission emerging in the soft band in most of the obscured spectra (Lanzuisi et al. 2015a).  
The normalization of this component is forced to be $<10\%$ of the primary component.

Once the best fit is obtained, we run a Monte Carlo Markov Chain (MCMC) within {\it Xspec} (v. 12.8.2), to evaluate
the \nh\ probability distribution function (PDF).
In this way we can collect a sample of CT AGN candidates, by counting
each source that is close to the CT boundary, only for the part of the PDF that exceeds this boundary.
Fig. 1 and 2 show the unfolded spectra and residuals for two CT AGN selected in this way. The bottom panels show
the \nh\ PDF (in red the part of the PDF that is taken into account to weight the source). 
Thanks to this analysis we were able to select a sample of 64 obscured sources, at $0.04<z<3.46$, 
having $>5\%$ probability of being CT (i.e. a fraction of them is not nominally CT).
Summing up only the fraction of the PDF of each source that is above $10^{24}$ \cm2,
we obtained a number of CT sources of N$_{\rm CT}=33.85$.

\begin{figure}[t]
\begin{center}
\includegraphics[width=7cm,height=5cm]{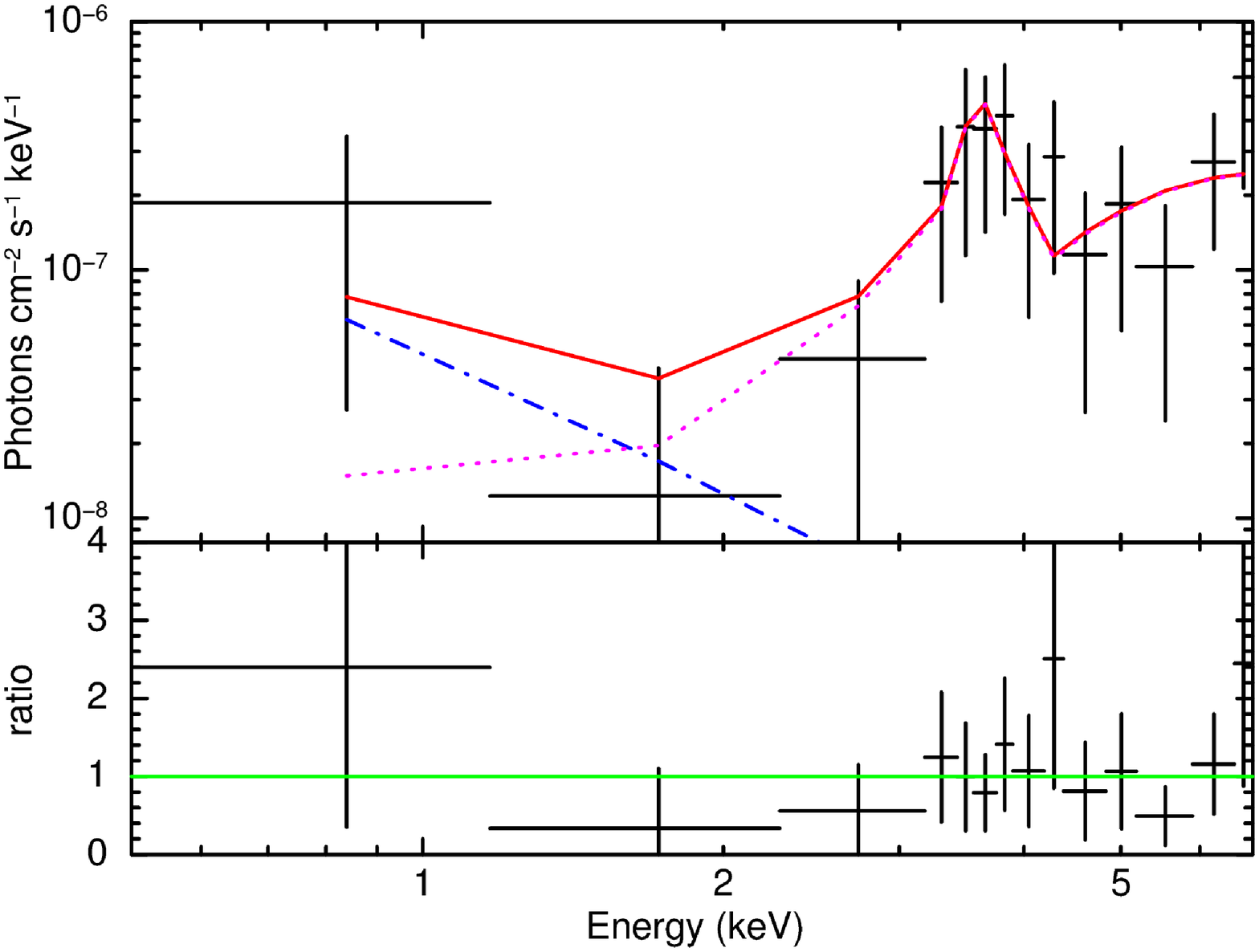}\vspace{0.1cm}
\includegraphics[width=7cm,height=5cm]{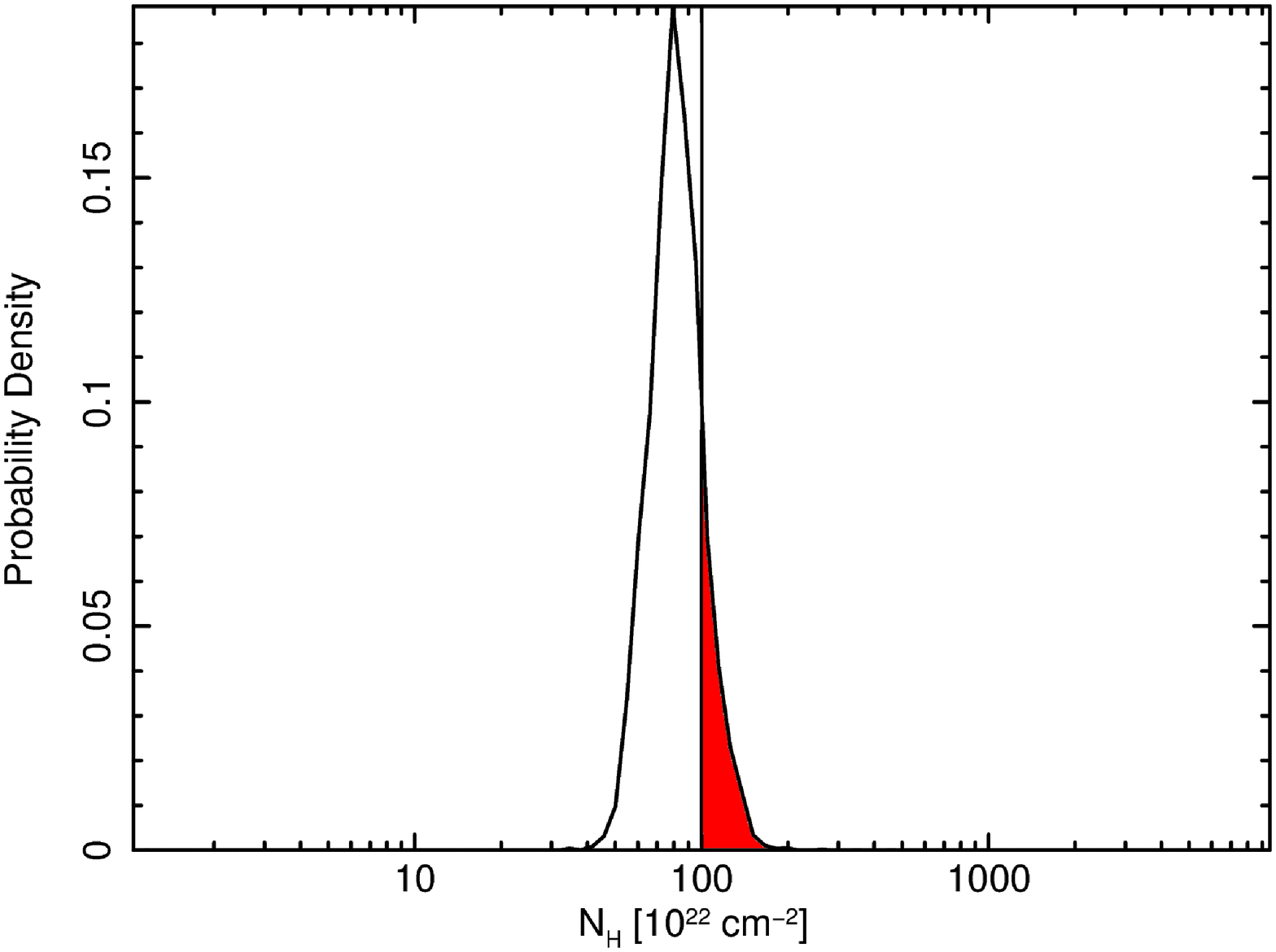}
\caption{{\small {\it Top:} Unfolded spectrum and data-to-model ratio of the CT candidate LID\_633 at $z=0.706$.
In magenta is shown the BNtorus component, in blue the soft powerlaw, and in red the total.
{\it Bottom:} PDF of \nh\ for the spectrum shown above.}}
\label{fig:spec1}
\end{center}
\end{figure}
\begin{figure}[t]
\begin{center}
\includegraphics[width=7cm,height=5cm]{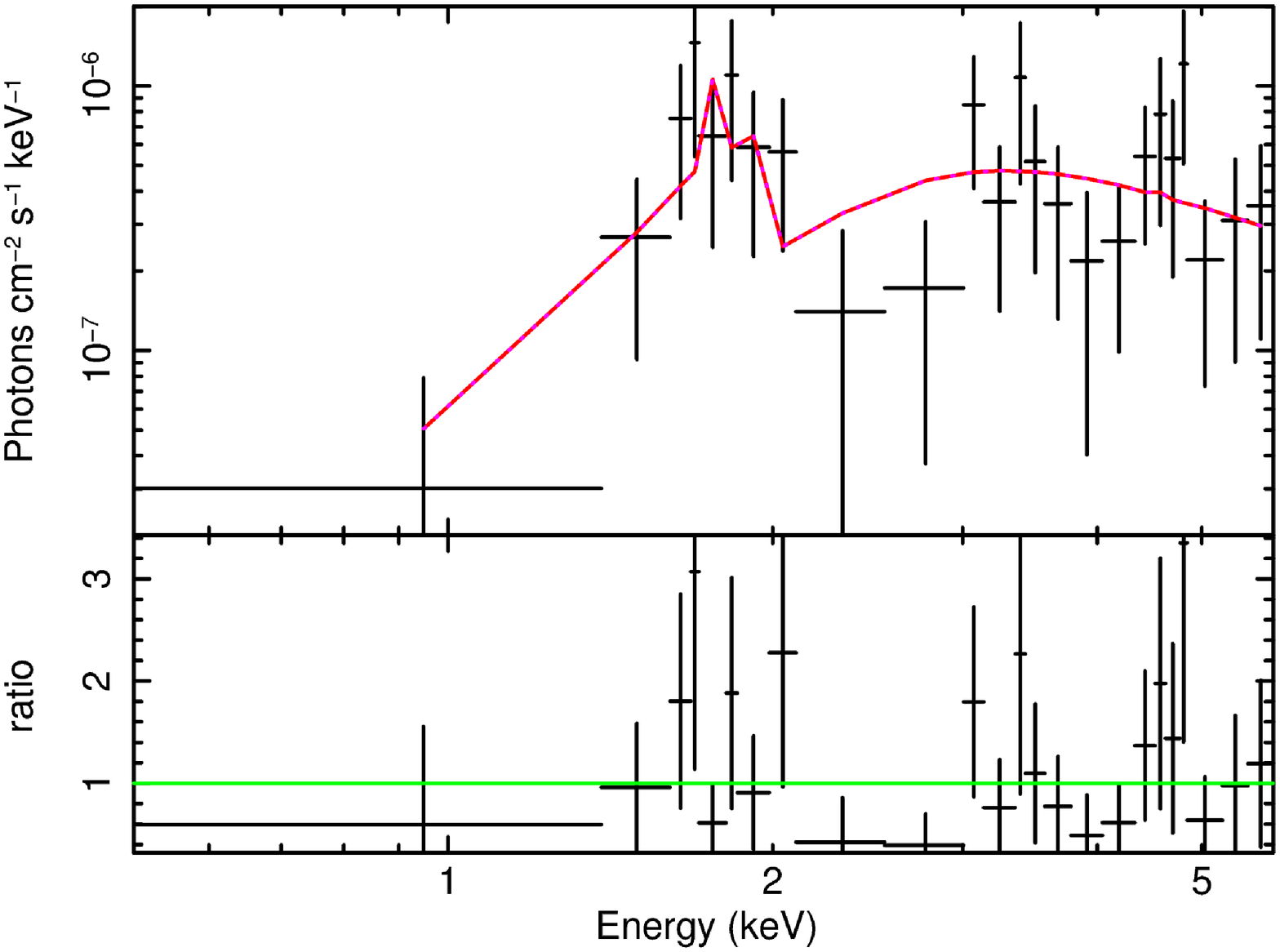}\vspace{0.1cm}
\includegraphics[width=7cm,height=5cm]{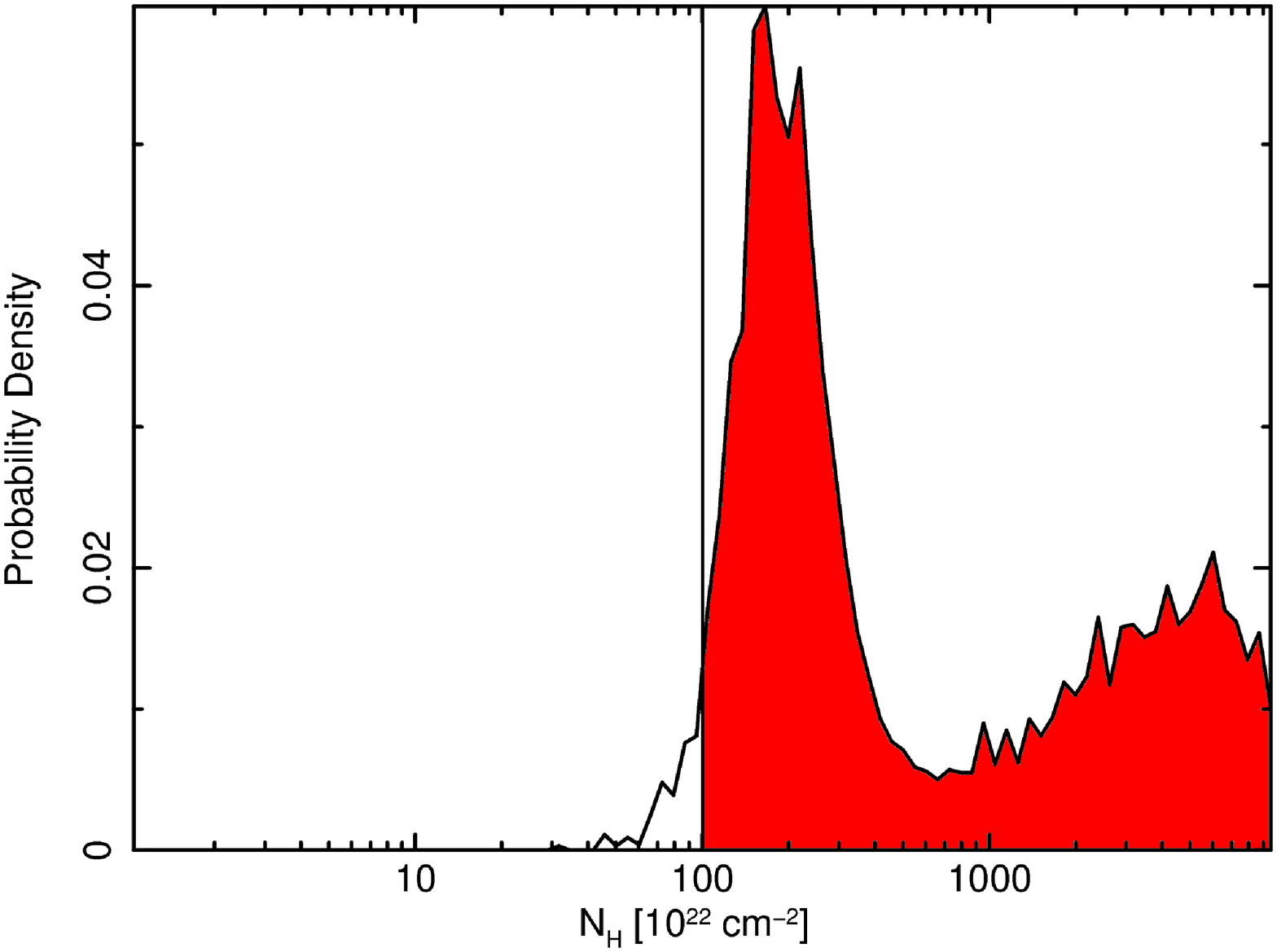}
\caption{{\small {\it Top:} Unfolded spectrum and data-to-model ratio of the CT candidate LID\_1002 at $z=2.612$.
This source only required the BNtorus component.
{\it Bottom:} PDF of \nh\ for the same source.}}
\label{fig:spec2}
\end{center}
\end{figure}

\section{Results}

\begin{figure}
\begin{center}
\includegraphics[width=7cm,height=6cm]{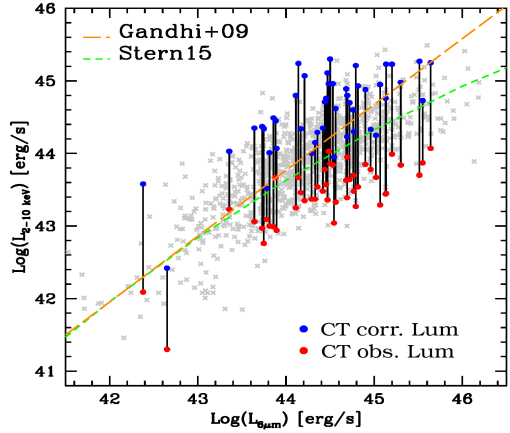}
\caption{{\small Distribution of the AGN Mid-IR ($6 \mu m$) luminosity, as derived from SED fitting, after subtracting the 
host star-formation emission, and the 2-10 keV luminosity, for the $\sim2000$ sources analyzed in Marchesi et al. 2016 (gray
crosses). Red circles show the observed \lum\ for the CT candidates, while blue circles show 
the absorption-corrected \lum. The orange (green) curve shows the relation published in 
Gandhi et al. 2009 (Stern 2015).
}}
\label{fig:lumlum}
\end{center}
\end{figure}

The sample of selected CT candidates spans a wide range in redshift, $0.04<z<3.46$ (50\% of them have
a photometric redshift) and absorption corrected X-ray luminosity, 
$43.5<$Log(\lum)$<45.8$ \ergs, 
with the exception of source LID\_1791 having Log(\lum)$=42.2$ \ergs at z=0.04, 
identified as a CT AGN in Civano et al. (2015)\footnote{This source is one of the two, with LID\_633, that is 
detected with \nus. The \nh\ values are consistent between our fit and the ones performed using also the 
hard X-ray data (Zappacosta et al. 2016 in prep.).}.

Given the large correction applied due to the obscuration, almost all the sources in the sample are in the quasar 
regime (\lum$>10^{44}$ \ergs, see blue points in fig. 3). In order to verify if these luminosities are reasonable, 
and therefore if our estimate of the obscuration is correct and not overestimated, 
we verified that the X-ray luminosity derived 
after correcting for the obscuration is consistent, within $\sim1$ dex
with the mid-IR AGN luminosity as computed from the spectral energy distribution (SED) 
fitting (either from Delvecchio et al. 2015 or Suh et al. 2016), after removing the host star-formation emission. 
Fig. 3 shows the distribution of the AGN Mid-IR ($6 \mu m$) luminosity, vs. 
the absorption-corrected \lum, for the $\sim2000$ sources analyzed in Marchesi et al. 2016 (gray crosses).
Red circles show the observed \lum\ of the CT candidates, while blue circles show 
the absorption-corrected \lum. 

Almost all our CT AGN have their \lmir-\lum\ within the typical scatter 
from the average relation (the orange or green curves, the \lmir-\lum\ relations published in 
Gandhi et al. 2009 and Stern 2015, respectively). 
However the vast majority lie in the upper part of the distribution: this is a clear selection effect (see e.g.
Lanzuisi et al. 2015a) due to the fact that CT sources with intrinsic \lum\ in the lower part of the diagram
would have not been detected in X-rays, having an absorbed \lum\ (and hence flux) below the detection threshold
of the survey. This already tells us that, despite our effort, a sizable fraction of CT AGN is still missing due to their 
low X-ray fluxes.

\subsection{LogN-LogS}

\begin{figure}
\begin{center}
\includegraphics[width=7cm,height=5.5cm]{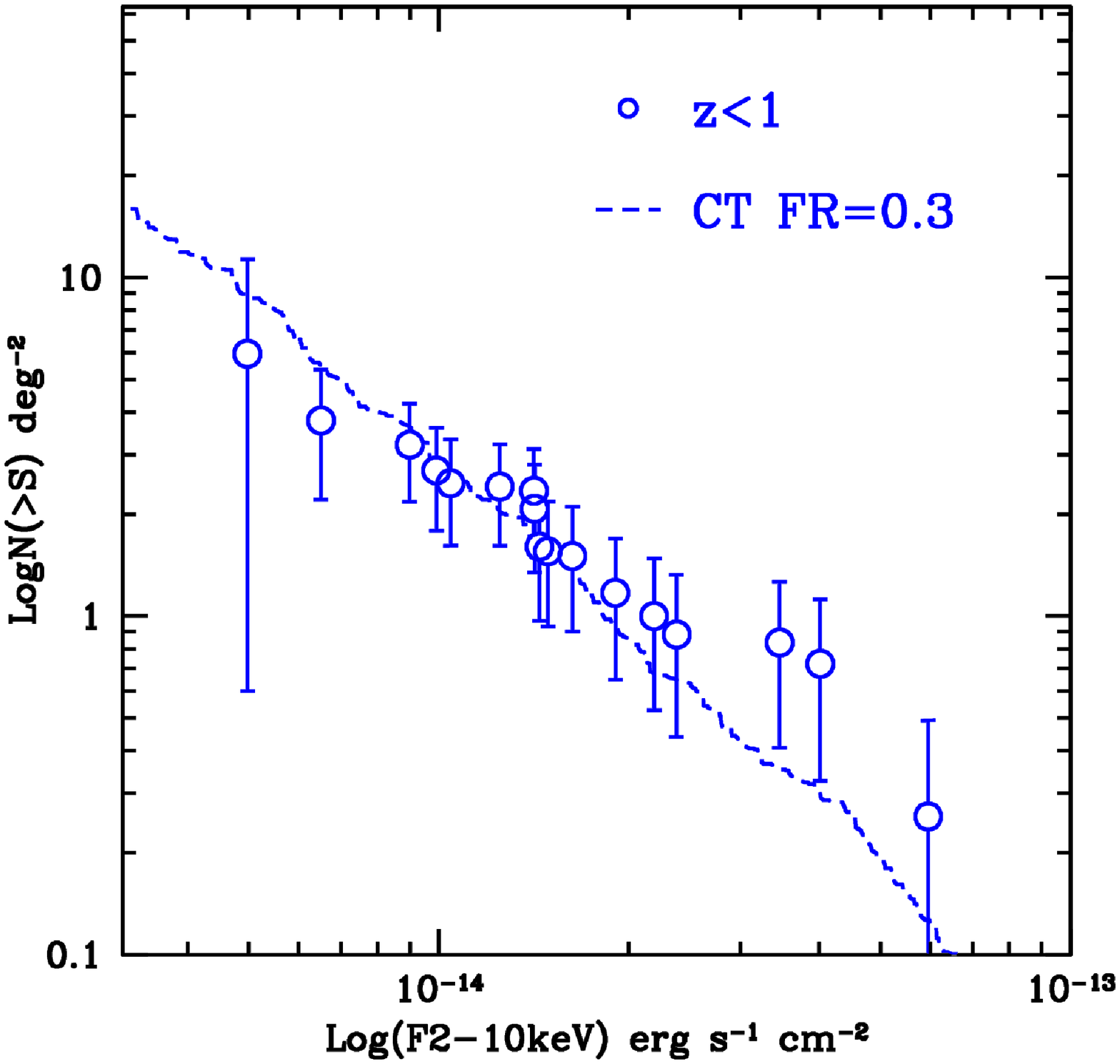}\vspace{0.1cm}
\includegraphics[width=7cm,height=5.5cm]{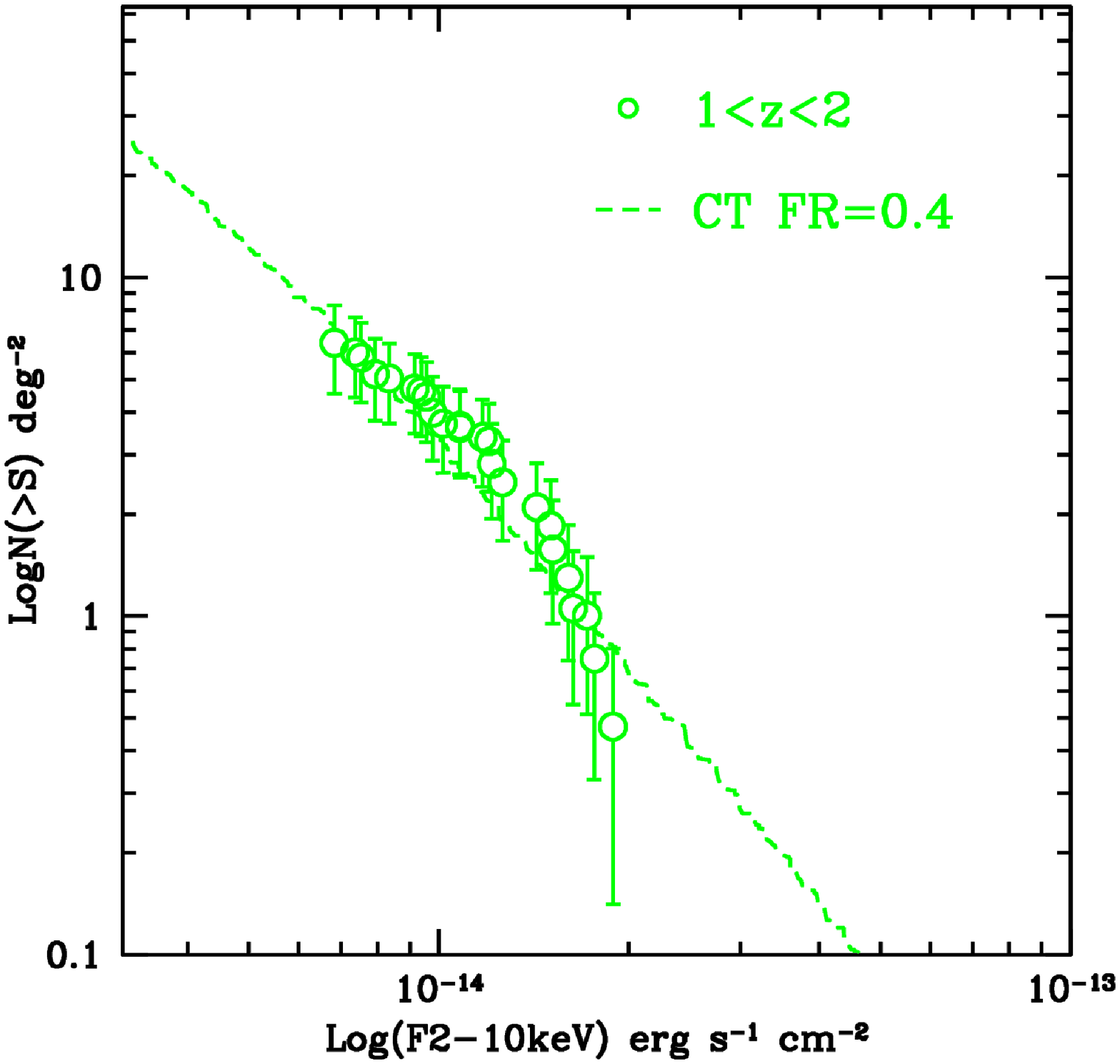}\vspace{0.1cm}
\includegraphics[width=7cm,height=5.5cm]{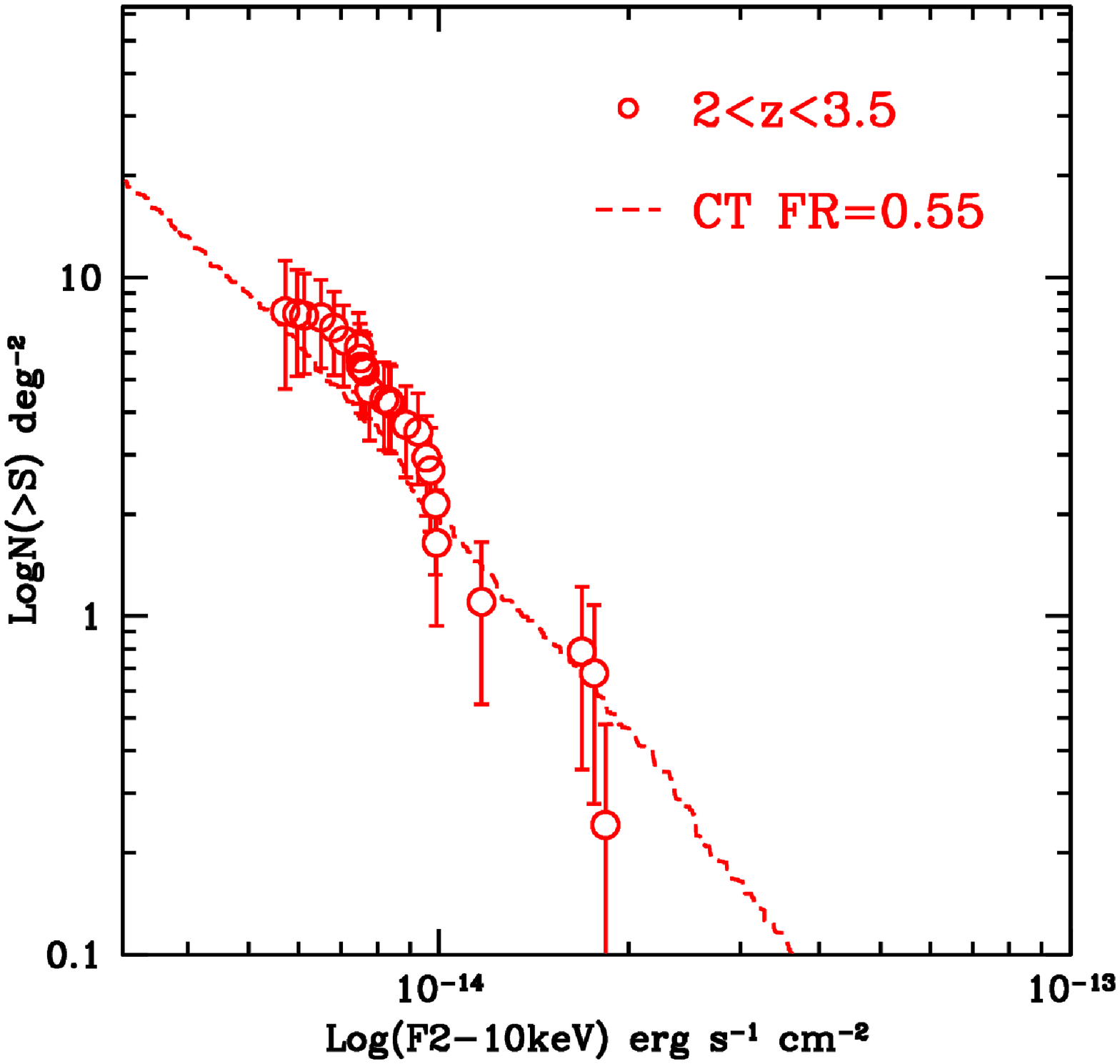}
\caption{{\small Observed LogN-LogS of CT candidates at $z<1$ (top), $1<z<2$ (center) and $2<z<3.5$ (bottom),
compared with the model for Akylas et al. 2012, with different CT fractions (FR).}}
\label{fig:lognlogs}
\end{center}
\end{figure}

To overcome this incompleteness, we can look at the LogN-LogS distribution,
where we can compare CXB models with the number of sources detected above the flux threshold of the survey.
To investigate a possible variation of the CT fraction with redshift,
we divided the sample into three redshift bins, and compared the observed LogN-LogS,
with the one predicted from the Akylas et al. (2012) model (Fig. 4).
This model assumes $\Gamma=1.9$, a high energy cut-off of 195 keV and a fraction of the reflected 
component flux with respect to the primary powerlaw of 0.05 in the 2-10 keV band.
To match the observed LogN-LogS of the CT AGN selected in COSMOS, 
the fraction of sources in the Log(\nh) bins 24-26 (equally distributed between the 24-25 and 25-26 bin)
must increase, from 30\% in the $z<1$ bin, to 40\% in the $1<z<2$, up to 55\% in the $2<z<3.5$ bin.
These fractions are systematicly higher than what predicted from other CXB models such as 
the one of Gilli et al. (2007) and Treister et al. (2009).

Brightman \& Ueda (2012) have already found a hint of an increase of the CT fraction
from $z<0.1$ (20\% derived in Burlon et al. 2011) to $z>1$ (40\% based on only 8 sources),
while other authors have found no evolution of the CT fraction at high z (e.g. Buchner et al. 2015).
We confirm the increase in the fraction of CT, especially at $2<z<3.5$, with a sample of 24 sources,
(the sum of all the probability distributions gives $\sim13$ sources).
This result is in agreement with the idea that at, high redshift, other factors,
such as the concentration of dense gas and dust clouds in the central region of the host galaxy
(possibly induced by gas-rich mergers), may contribute to CT obscuration, on top of the canonical molecular torus
(e.g. Hopkins et al. 2006).

\subsection{Extremely obscured CT}

Of particular interest is the observed distribution of \nh\ above $10^{25}$ cm$^{-2}$, first of all because 
it is mostly unknown, even at low redshift (see e.g. Maiolino et al. 2003), 
and second because it is not possible to derive any estimate on the number
of heavily CT sources with the indirect method of the CXB synthesis models.

Very few (e.g. Piconcelli eta l. 2011, Gandhi et al. 2013, or the debated case of Arp220, Wilson et al. 2014) 
of these reflection dominated sources is known in the local universe 
(see the compilation of Gandhi et al. 2014), and even more uncertain is their fraction
above z$\sim0.1$.

In Lanzuisi et al. (2015b) we have found the first \nh$=10^{25}$ \cm2\ CT AGN candidate in COSMOS, thanks to 
several multiwavelength indicators ([NeV], IR and bolometric luminosity from SED). 
With the analysis presented here we were able to find, only through X-ray spectral analysis,
nine sources that have the \nh\ PDF completely above $10^{24}$ \cm2\ and the upper boundary pegged at the upper 
limit of the model adopted $N_H(up)=10^{26}$ \cm2. 
Fig. 5 show the spectra and PDF of two such reflection dominated sources, where a very strong Fe K$\alpha$
line is superimposed to a flat continuum.
Most (7/9) of these sources have $z\simgt1.5$, because above these values both the Fe line and the Compton hump
are easier to detect.

\begin{figure}[t]
\begin{center}
\includegraphics[width=7cm,height=5cm]{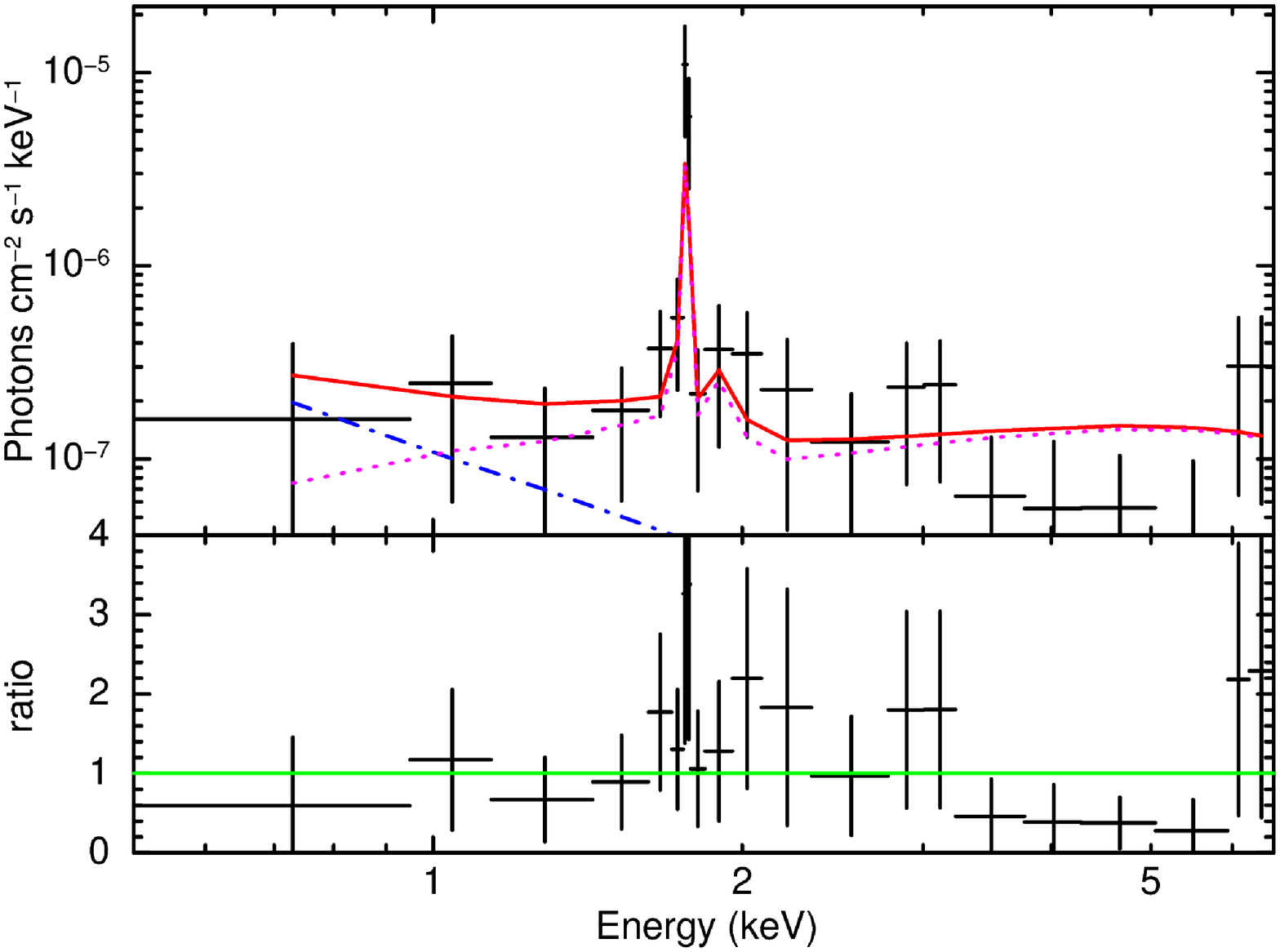}\vspace{0.1cm}
\includegraphics[width=7cm,height=5cm]{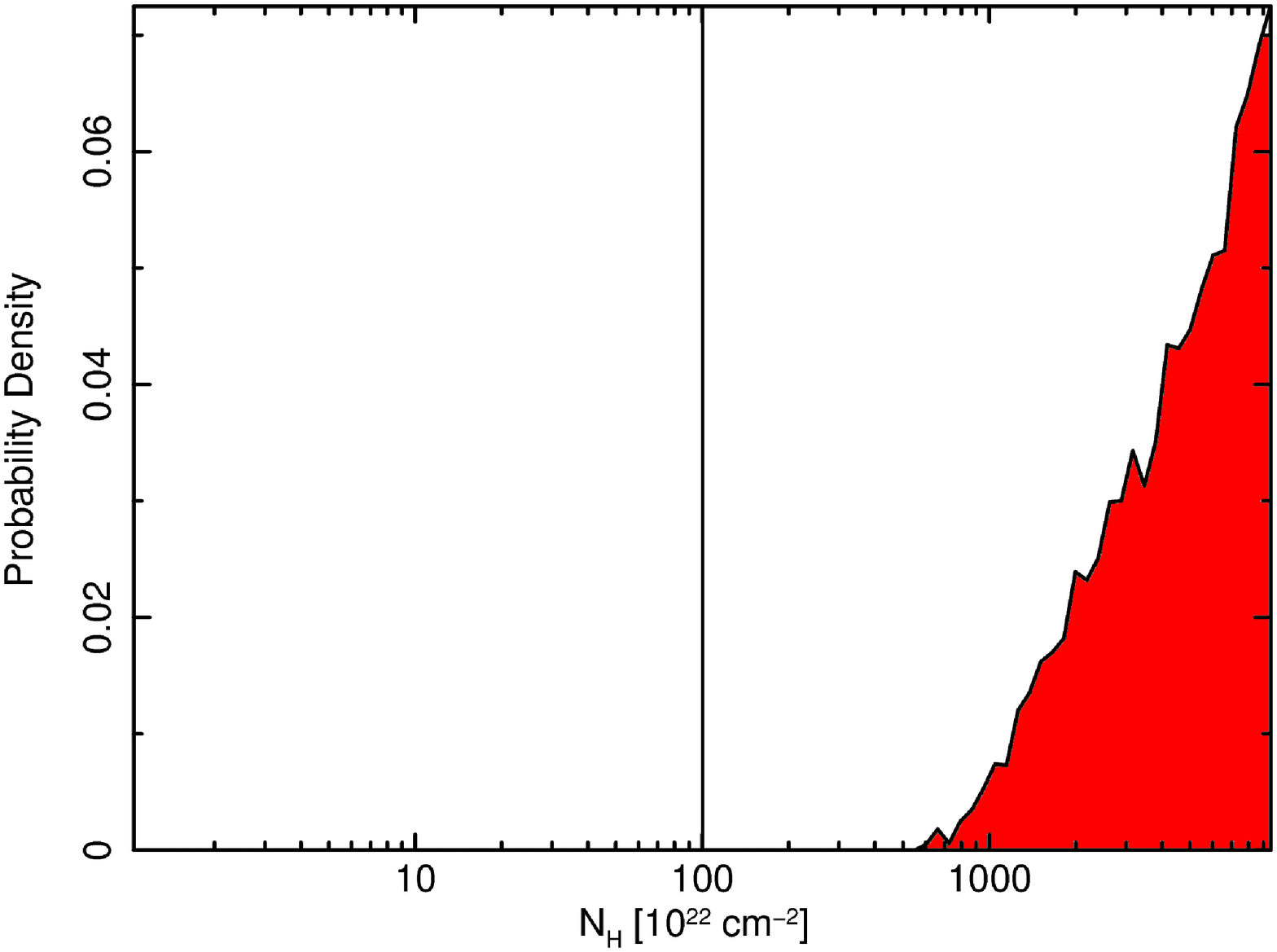}
\caption{{\small {\it Top:} Unfolded spectrum and data-to-model ratio of the heavily CT candidate CID\_708.
{\it Bottom:} PDF of \nh\ for the same source.}}
\label{fig:spec3}
\end{center}
\end{figure}

\begin{figure}[t]
\begin{center}
\includegraphics[width=7cm,height=5cm]{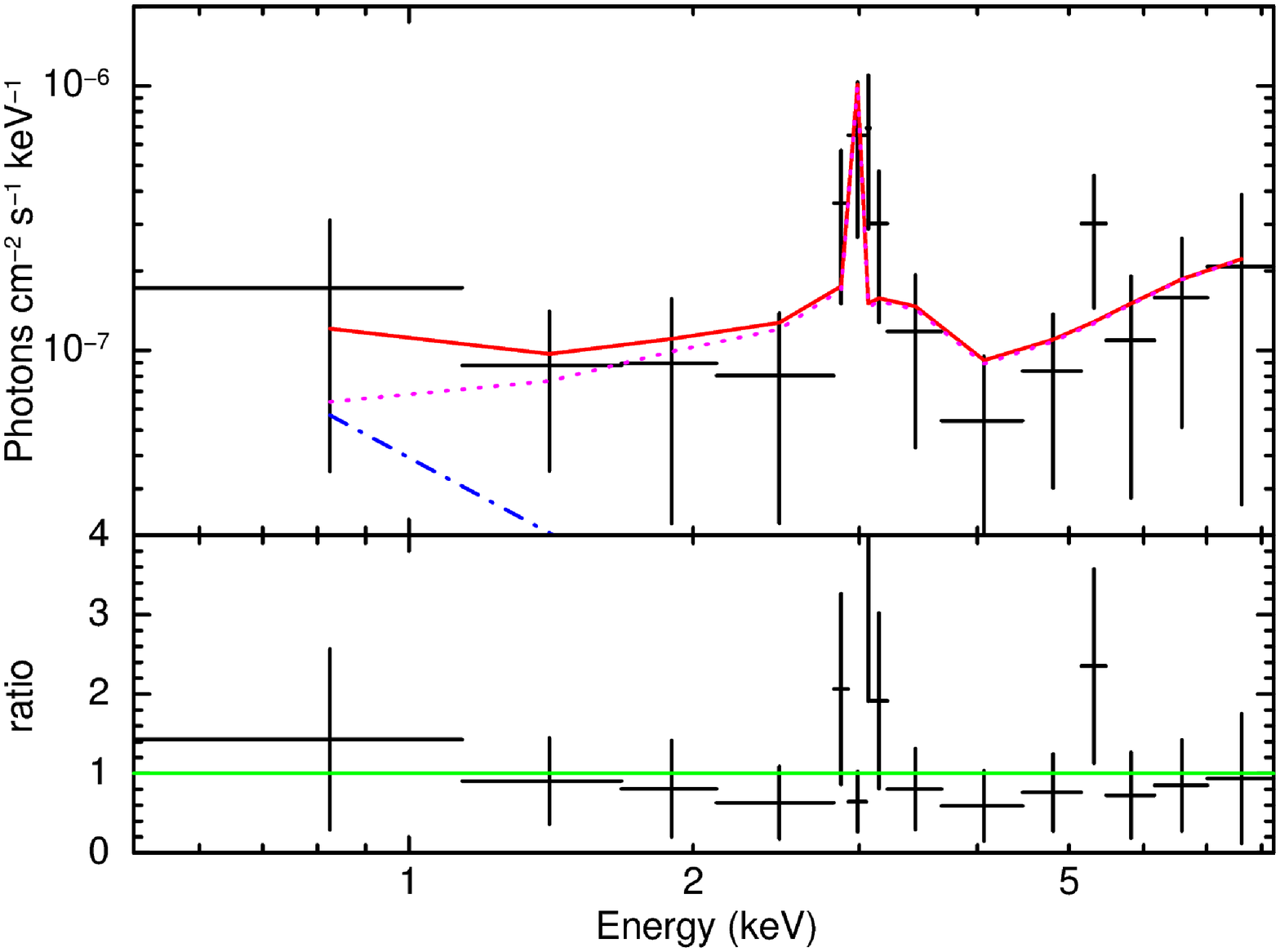}\vspace{0.1cm}
\includegraphics[width=7cm,height=5cm]{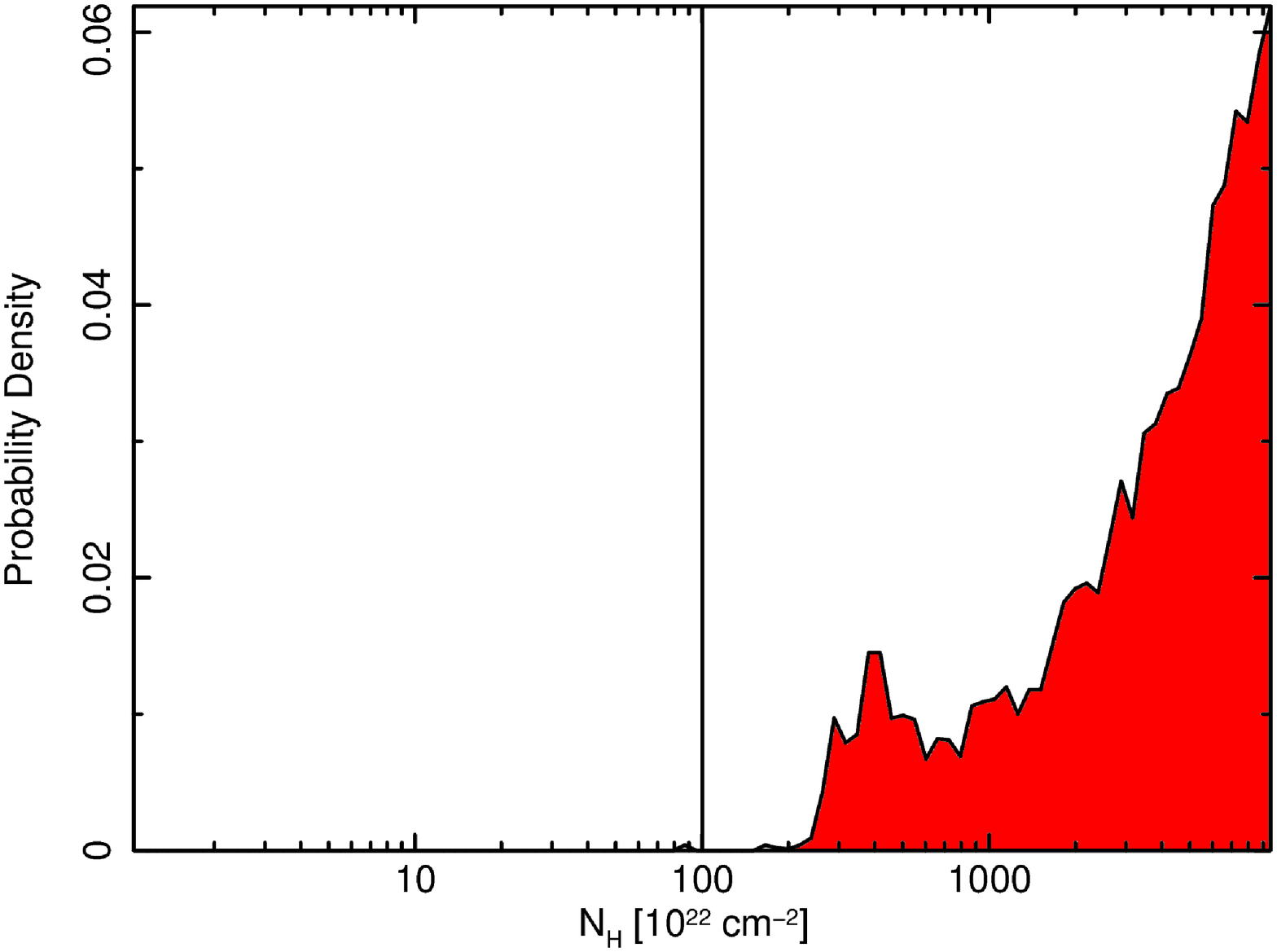}
\caption{{\small As fig. 5, for source LID\_390.}
}
\label{fig:spec4}
\end{center}
\end{figure}

\section{Conclusions}

Thanks to the combination of new dedicated models and sophisticated analysis techniques, specifically 
developed for low counts statistics, recent works have found a growing number of CT AGN 
at high redshift in deep X-ray surveys (e.g. Brightman et al. 2014, Buchner et al. 2015). 
We applied a similar approach to the large catalog of X-ray detected AGN in the COSMOS-Legacy 
survey and demonstrated that not only are mildly CT sources up to $z\sim3.5$ recoverable via X-ray spectral analysis,
but also reflection dominated, heavily CT sources at $z\sim2$ can be efficiently identified.

We underline, however, that X-rays need to be complemented by rich multiwavelength data, in order 
to verify that the solution at high \nh\ - and therefore high \lum\ - found for each of these sources,
is not only statistically motivated, given the available X-ray spectra, but also physically reasonable, 
given all the other multiwavelength data available. 

\subsection{Implications for XMM-Newton}

While \chandra\ is more efficient in detecting faint (high z and/or obscured) sources\footnote{by about a 
factor $\sim2$, as can be derived comparing the total exposure time needed, and the number of CT AGN selected,
in the \chandra\ (this work) and \xmm\ (Lanzuisi et al. 2015a) catalogs of a medium/deep survey like COSMOS},
the spectra obtained with long \xmm\ exposures have a much higher number of net counts (if the source is bright enough, 
i.e. few$\times10^{-15}$ \cgs), and they allow to constraint simultaneously the 
column density, even above $10^{25}$ cm$^{-2}$ and the reflection fraction, two crucial 
parameters for CXB models. 

As an example we show here the \xmm\ spectrum of source XID-202 in the XMM-CDFS catalog (Ranalli et al. 2013).
The nominal 3.4 Ms exposure time invested by \xmm\ (corresponding to an effective exposure time of $\sim6$ Ms
between pn and MOS1+2) allowed to collect $\sim1500$ total net counts (in Fig. 7
we show only the pn spectrum for clarity) for this z=3.7 CT source (Comastri et al. 2011).
The spectral quality is high enough to allow us to use a more complex CT model (MYTorus, in the decoupled version) and 
constraint simultaneously the \nh\ and the reflection fraction (Fig 7 bottom panel, thick lines).
On the other hand, the nominal 4Ms collected with \chandra\ (3.6 effective exposure at the source position)
give only 380 net counts. Despite the much lower background, the uncertainties on these parameters, 
obtained from the \chandra\ spectrum are much larger (dashed lines).

\begin{figure}
\begin{center}
\includegraphics[width=7cm,height=5cm]{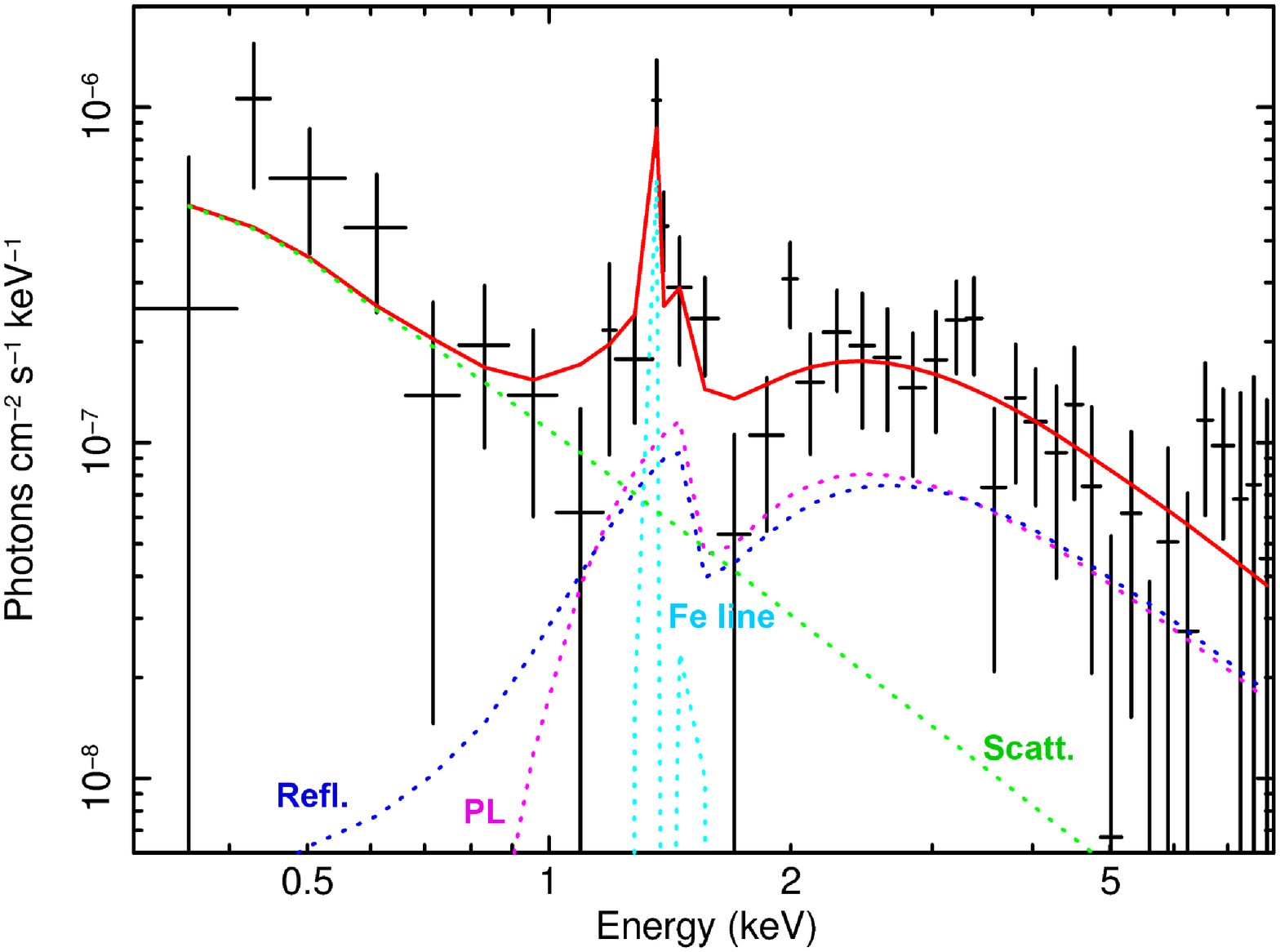}
\vspace{0.5cm}
\includegraphics[width=7cm,height=5cm]{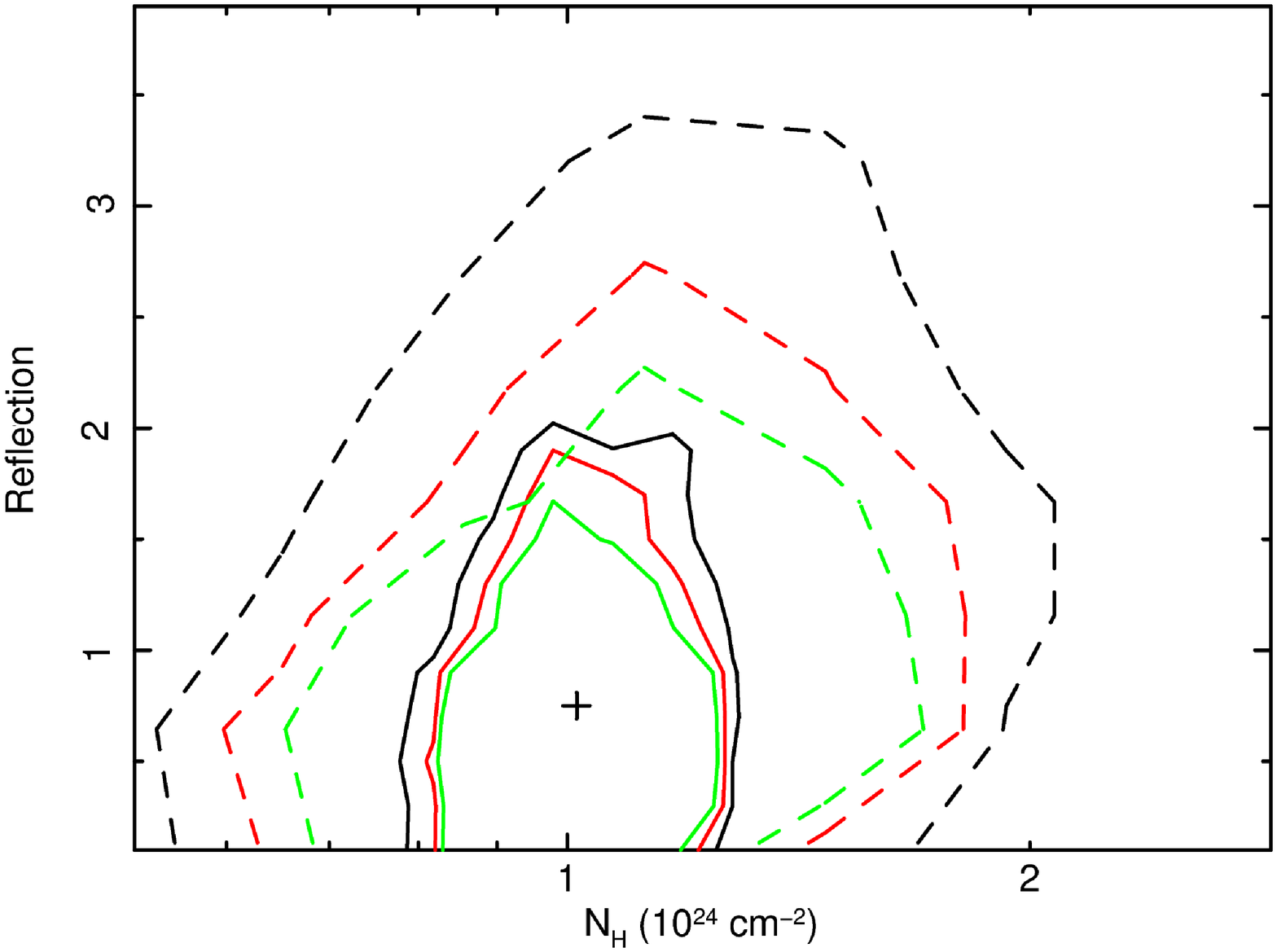}
\caption{{\small {\it Top:} Unfolded spectrum of source XID202 from the XMM-CDFS catalog, a CT AGN at z=3.7.
The different components (primary powerlaw, reflection, Fe line complex, scattered emission) 
are labeled with different colors.  
{\it Bottom:} 99, 90 and 68\% confidence contours obtained from the \xmm\ spectrum (thick lines) and from the \chandra\
one (dotted lines), for the \nh\ and reflection fraction.}}
\label{fig:cdfs}
\end{center}
\end{figure}

Of course sources of similar intrinsic luminosity (\lum$>5\times10^{44}$ \ergs) at high redshift are rare 
(e.g. Vito et al. 2014). Therefore, the most efficient way to collect a decent sample ( e.g. 10 sources) 
of high redshift CT AGN, at these flux levels (F$_{2-10}\sim5\times10^{-15}$ \cgs), 
for which all the different parameters of a complex CT model can be determined, may be to spend long (e.g. 1Ms) 
exposure times over CT candidates at high z selected in existing/ongoing large area surveys,
such as XMM-XXL (Pierre et al. 2015) or S82 (La Massa et al. 2013).

\acknowledgements

The author acknowledges financial support from the CIG grant ``eEASY'' n. 321913 and from ASI-INAF grant n. 2014-045-R.0.


\begin{thebibliography}{}


\bibitem[Akylas et al.(2012)]{2012A&A...546A..98A} Akylas, A., Georgakakis, A., Georgantopoulos, I, et al.\ 2012, \aap, 546, A98 
\bibitem[Ballantyne et al.(2011)]{2011ApJ...736...56B} Ballantyne, D.~R., Draper, A.~R., Madsen, K.~K., et al.\ 2011, \apj, 736, 56 
\bibitem[Brightman \& Nandra(2011)]{2011MNRAS.413.1206B} Brightman, M., \& Nandra, K.\ 2011, \mnras, 413, 1206 
\bibitem[Brightman \& Ueda(2012)]{2012MNRAS.423..702B} Brightman, M., \& Ueda, Y.\ 2012, \mnras, 423, 702 
\bibitem[Brightman et al.(2014)]{2014arXiv1406.4502B} Brightman, M., Nandra, K., Salvato, M., et al.\ 2014, arXiv:1406.4502 
\bibitem[Burlon et al.(2011)]{2011ApJ...728...58B} Burlon, D., Ajello, M., Greiner, J., et al.\ 2011, \apj, 728, 58 
\bibitem[Buchner et al.(2015)]{2015ApJ...802...89B} Buchner, J., Georgakakis, A., Nandra, K., et al.\ 2015, \apj, 802, 89 
\bibitem[Cappelluti et al.(2009)]{2009A&A...497..635C} Cappelluti, N., Brusa, M., Hasinger, G., et al.\ 2009, \aap, 497, 635 
\bibitem[Castell{\'o}-Mor et al.(2013)]{2013A&A...556A.114C} Castell{\'o}-Mor, N., Carrera, F.~J., Alonso-Herrero, A., et al.\ 2013, \aap, 556, A114 
\bibitem[Civano et al.(2012)]{2012ApJS..201...30C} Civano, F., Elvis, M., Brusa, M., et al.\ 2012, \apjs, 201, 30 
\bibitem[Civano et al.(2015)]{2015ApJ...808..185C} Civano, F., Hickox, R.~C., Puccetti, S., et al.\ 2015, \apj, 808, 185 
\bibitem[Civano et al.(2016)]{2016ApJ...819...62C} Civano, F., Marchesi, S., Comastri, A., et al.\ 2016, \apj, 819, 62 
\bibitem[Comastri et al.(1995)]{1995A&A...296....1C} Comastri, A., Setti, G., Zamorani, G., \& Hasinger, G.\ 1995, \aap, 296, 1 
\bibitem[Comastri et al.(2011)]{2011A&A...526L...9C} Comastri, A., Ranalli, P., Iwasawa, K., et al.\ 2011, \aap, 526, L9 
\bibitem[Comastri et al.(2015)]{2015A&A...574L..10C} Comastri, A., Gilli, R., Marconi, A., et al.\ 2015, \aap, 574, L10 
\bibitem[Delvecchio et al.(2015)]{2015MNRAS.449..373D} Delvecchio, I., Lutz, D., Berta, S., et al.\ 2015, \mnras, 449, 373 
\bibitem[Elvis et al.(2009)]{2009ApJS..184..158E} Elvis, M., Civano, F., Vignali, C., et al.\ 2009, \apjs, 184, 158 
\bibitem[Fiore et al.(2008)]{2008ApJ...672...94F} Fiore, F., Grazian, A., Santini, P., et al.\ 2008, \apj, 672, 94-101 
\bibitem[Gandhi et al.(2009)]{2009A&A...502..457G} Gandhi, P., Horst, H., Smette, A., et al.\ 2009, \aap, 502, 457 
\bibitem[Georgantopoulos et al.(2011)]{2011A&A...531A.116G} Georgantopoulos, I., Dasyra, K.~M., Rovilos, E., et al.\ 2011, \aap, 531, A116 
\bibitem[Georgantopoulos et al.(2013)]{2013A&A...555A..43G} Georgantopoulos, I., Comastri, A., Vignali, C., et al.\ 2013, \aap, 555, A43 
\bibitem[Gilli et al.(2007)]{2007A&A...463...79G} Gilli, R., Comastri, A., \& Hasinger, G.\ 2007, \aap, 463, 79 
\bibitem[Gilli et al.(2010)]{2010A&A...519A..92G} Gilli, R., Vignali, C., Mignoli, M., et al.\ 2010, \aap, 519, AA92 
\bibitem[Hasinger et al.(2007)]{2007ApJS..172...29H} Hasinger, G., Cappelluti, N., Brunner, H., et al.\ 2007, \apjs, 172, 29 
\bibitem[Houck et al.(2005)]{2005ApJ...622L.105H} Houck, J.~R., Soifer, B.~T., Weedman, D., et al.\ 2005, \apjl, 622, L105 
\bibitem[Ilbert et al.(2009)]{2009ApJ...690.1236I} Ilbert, O., Capak, P., Salvato, M., et al.\ 2009, \apj, 690, 1236 
\bibitem[Kormendy \& Ho(2013)]{2013ARA&A..51..511K} Kormendy, J., \& Ho, L.~C.\ 2013, \araa, 51, 511 
\bibitem[Lacy et al.(2004)]{Lacy2004} Lacy, M., Storrie-Lombardi, L. J., Sajina, A., et al., 2004, ApJS, 154, 166
\bibitem[Lanzuisi et al.(2009)]{2009A&A...498...67L} Lanzuisi, G., Piconcelli, E., Fiore, F., et al.\ 2009, \aap, 498, 67 
\bibitem[Lanzuisi et al.(2013)]{2013MNRAS.431..978L} Lanzuisi, G., Civano, F., Elvis, M., et al.\ 2013, \mnras, 431, 978 
\bibitem[Lanzuisi et al.(2015)]{2015A&A...573A.137L} Lanzuisi, G., Ranalli, P., Georgantopoulos, I., et al.\ 2015a, \aap, 573, AA137 (L15)
\bibitem[Lanzuisi et al.(2015)]{2015A&A...578A.120L} Lanzuisi, G., Perna, M., Delvecchio, I., et al.\ 2015b, \aap, 578, A120 
\bibitem[Marchesi et al.(2016)]{2016ApJ...817...34M} Marchesi, S., Civano, F., Elvis, M., et al.\ 2016b, \apj, 817, 34 
\bibitem[Marchesi et al.(2016)]{2016arXiv160805149M} Marchesi, S., Lanzuisi, G., Civano, F., et al.\ 2016a, arXiv:1608.05149 
\bibitem[Marconi et al.(2004)]{2004MNRAS.351..169M} Marconi, A., Risaliti, G., Gilli, R., et al.\ 2004, \mnras, 351, 169 
\bibitem[Mart\'{i}nez-Sansigre et al.(2005)]{MartinezSansigre2005} Mart\'{i}nez-Sansigre, A., Rawlings, S., Lacy, M., et al., 2005, Nature, 436, 666
\bibitem[Murphy \& Yaqoob(2009)]{2009MNRAS.397.1549M} Murphy, K.~D., \& Yaqoob, T.\ 2009, \mnras, 397, 1549 
\bibitem[Mignoli et al.(2013)]{2013A&A...556A..29M} Mignoli, M., Vignali, C., Gilli, R., et al.\ 2013, \aap, 556, A29 
\bibitem[Ranalli et al.(2013)]{2013A&A...555A..42R} Ranalli, P., Comastri, A., Vignali, C., et al.\ 2013, \aap, 555, A42 
\bibitem[Risaliti et al.(1999)]{1999ApJ...522..157R} Risaliti, G., Maiolino, R., \& Salvati, M.\ 1999, \apj, 522, 157 
\bibitem[Salvato et al.(2011)]{2011ApJ...742...61S} Salvato, M., Ilbert, O., Hasinger, G., et al.\ 2011, \apj, 742, 61 
\bibitem[Scoville et al.(2007)]{2007ApJS..172....1S} Scoville, N., Aussel, H., Brusa, M., et al.\ 2007, \apjs, 172, 1 
\bibitem[Soltan(1982)]{1982MNRAS.200..115S} Soltan, A.\ 1982, \mnras, 200, 115 
\bibitem[Stern(2015)]{2015ApJ...807..129S} Stern, D.\ 2015, \apj, 807, 129 
\bibitem[Tozzi et al.(2006)]{2006A&A...451..457T} Tozzi, P., Gilli, R., Mainieri, V., et al.\ 2006, \aap, 451, 457 
\bibitem[Treister et al.(2009)]{2009ApJ...696..110T} Treister, E., Urry, C.~M., \& Virani, S.\ 2009, \apj, 696, 110 
\bibitem[Ueda et al.(2014)]{2014ApJ...786..104U} Ueda, Y., Akiyama, M., Hasinger, G., et al.\ 2014, \apj, 786, 104 
\bibitem[Vignali et al.(2006)]{2006MNRAS.373..321V} Vignali, C., Alexander, D.~M., \& Comastri, A.\ 2006, \mnras, 373, 321 
\bibitem[Vignali et al.(2014)]{2014A&A...571A..34V} Vignali, C., Mignoli, M., Gilli, R., et al.\ 2014, \aap, 571, A34 

  
 \end{thebibliography}
\end{document}